\documentclass[preprint,showpacs,preprintnumbers,amsmath,amssymb]{revtex4}
\usepackage{graphicx}
\usepackage{dcolumn}
\usepackage{bm}
\usepackage{color}
\usepackage[applemac]{inputenc}

\begin{document}
\title{Spectral dependence of purely-Kerr driven filamentation in air and argon}

\author{W. Ettoumi$^{1}$}
\author{P. B\'ejot$^{2}$}
\author{Y. Petit$^1$}
\author{V. Loriot$^{2,3}$}
\author{E. Hertz$^{2}$}
\author{O. Faucher$^{2}$}
\author{B. Lavorel$^{2}$}
\author{J. Kasparian$^{1}$}\email{jerome.kasparian@unige.ch}
\author{J.-P. Wolf$^{1}$}
\affiliation{(1) Universit\'e de Gen\`eve, GAP-Biophotonics, 20 rue
de l'Ecole de M\'edecine, 1211 Geneva 4, Switzerland}
\affiliation{(2) Laboratoire Interdisciplinaire Carnot de Bourgogne (ICB), UMR 5209 CNRS-Universit\'e de Bourgogne, 9 Av. A. Savary, BP 47 870, F-21078 DIJON Cedex, FRANCE}
\affiliation{(3) Instituto de Qu\'imica F\'isica Rocasolano, CSIC, C/Serrano, 119, 28006 Madrid, Spain}

\newlength{\textlarg}
\newcommand{\strike}[1]{%
   \settowidth{\textlarg}{#1}
   #1\hspace{-\textlarg}\rule[0.5ex]{\textlarg}{0.5pt}}

\begin{abstract}
Based on numerical simulations, we show that higher-order nonlinear indices (up to $n_8$ and $n_{10}$, respectively) of air and argon have a dominant contribution to both focusing and defocusing in the self-guiding of ultrashort laser pulses over most of the spectrum. Plasma generation and filamentation are therefore decoupled. As a consequence, ultraviolet wavelength may not be the optimal wavelengths for applications requiring to maximize ionization.
\end{abstract}

\maketitle

The filamentation of high-power, ultrashort laser pulses is generally described as a dynamic balance between Kerr self-focusing and the defocusing by the plasma generated at the non-linear focus \cite{BraunKLDSM95,ChinHLLTABKKS05,BergeSNKW07,CouaironM07,KasparianW08}. In this description, the Kerr-induced change in the refractive index is truncated to the first term $n_2I$, where 
$I$ is the local intensity. However, at higher intensities, the development has to be extended to higher-order terms in $I$, so that the real part of the refractive index at any frequency $\omega$ writes:
\begin{equation}
n(\omega)=n_0(\omega)+\Delta\text{$n$}_\text{Kerr}(\omega) = n_0(\omega)+\sum_{j\ge1}n_{2j}(\omega)I^j
\label{n_i}
\end{equation}
where the $n_{2j}(\omega)$ coefficients are related to the $(2j+1)^{th}$ order susceptibility tensor $\chi^{(2j+1)}(\omega)$, in the degenerate case where all considered fields are at frequency $\omega$:

\begin{equation}
n_{2j}(\omega)= \frac{(2j+1)!}{2^{j+1}j!(j+1)!}\frac{1}{\left(n_0^2(\omega)\epsilon_0 c\right)^j}  \Re e\left(\chi^{(2j+1)}(\omega) \right)
\label{n_2j_Chi_2jp1}
\end{equation}

In the last years, several numerical works have investigated the influence of the quintic nonlinear response on the filamentation dynamics at a wavelength of 800 nm, although without knowledge of its actual value \cite{AkozbekSBC01,VincotteB04,FibichI04,BergeSMKYFSW05, ZhangTWDW10}. They suggested that this term was defocusing, but considered it as marginal. Recently, the measurement of the higher-order Kerr indices at a wavelength of 800 nm up to $n_8$ in air and $n_{10}$ in argon \cite{LoriotHFL09} showed that they have alternate signs, and are therefore alternatively focusing and defocusing. 
Furthermore, by implementing these terms in a numerical simulation of filamentation, we have recently shown that the defocusing terms $n_4$ and $n_8$, rather than the plasma, provide the main regularizing process in the filamentation of 30 fs pulses in air at 800~nm, so that plasma generation and propagation equations are almost decoupled \cite{LoriotBHFLHKW09}. This finding provided an explanation to measurements of plasma-free filamentation \cite{MechainCADFPTMS04, DubietisGTT04} and predicts symmetrical temporal pulse shapes, in contrast with a balance between the instantaneous Kerr term and the time-integrated plasma contribution, which implies strongly asymmetric pulse shapes \cite {StibenzZS06}. It also explained the discrepancy, by almost two orders of magnitude, between experimentally measured and numerically derived electron densities within filaments. While the experiments yield some $10^{14}$ cm$^{-3}$ \cite{KasparianSC00,ThebergeLSBC06}, the numerical simulations require a few 10$^{16}$ cm$^{-3}$ to balance the $n_2I$ Kerr self-focusing term \cite{BergeSNKW07,CouaironM07}. The consideration of the higher-order Kerr terms also turned out to be necessary to obtain a quantitative agreement between numerical simulations and experimental results about the propagation of ultrashort infrared pulses in an argon-filled hollow-core fiber \cite{SchmidtBGSTBKWVKCL10,BejotSKWL10}.

However, the ionization rates are higher at shorter wavelengths, so that ionization is generally believed to be much stronger in the case of ultraviolet filamentation \cite{SchwartzRD00}, while its low efficiency in the infrared would expectedly prevent self-guiding of filaments. But the recently derived generalized Miller formul\ae \ \cite{EttoumiPKW10} predict larger absolute values of the higher-order non-linear indices at shorter wavelengths, with faster spectral dependencies for the higher orders. As a consequence, the relative contribution to self-guiding cannot be easily extrapolated from qualitative discussions. 


In this Letter, we numerically investigate these relative contributions from the near-ultraviolet to the near-infrared. Using the values of the $n_{2j}$ indices at any wavelength as obtained from the generalized Miller formul\ae \ \cite{EttoumiPKW10}, and the values recently measured at 800~nm \cite{LoriotHFL09}, we simulate the filamentation of laser pulses from 300~nm to 1600~nm in air and argon. We show that filaments are efficiently generated at all wavelengths, even in the infrared. Moreover, the plasma marginally contributes to self-guiding in the filamentation of 30 fs laser pulses in argon for wavelengths longer than typically 400 nm. In air, where the ionization of oxygen is about 100 times more efficient than in argon, the Kerr terms still provide the largest defocusing contribution, although the plasma contribution can be neglected over the whole propagation length only in the infrared. This finding strongly impacts the plasma generation. We also find that ultraviolet wavelengths are not optimal to maximize ionization in a filament.

Numerical simulations are performed, as described in detail in \cite{LoriotBHFLHKW09}, by solving a non-linear Schr\"odinger equation taking the higher-order Kerr terms into account. The reduced scalar envelope $\varepsilon$ defined such that $|\varepsilon(r,z,t)|^2=I(r,z,t)$, $I$ being the intensity, is assumed to vary slowly in time and along $z$: 

\begin{eqnarray}
\begin{aligned}
\label{NLSE}
&\partial_z\varepsilon =\frac{i}{2k}\triangle_{\bot}\varepsilon-i\frac{k''}{2}\partial_t^2\varepsilon \\
&+\left(1+\frac{i}{\omega}\frac{\partial}{\partial t}\right)i\frac{k}{n_0}\left(\sum_{j\ge1}{n_{2j}|\varepsilon|^{2j}}\right)\varepsilon
-i\frac{k}{2n_0^2\rho_c}\rho\varepsilon \\
&-\frac{\varepsilon}{2}\sum_{\ell=O_2,N_2}{\left(\sigma_\ell(\omega)\rho+\frac{W_\ell(|\varepsilon|^2,\omega)U_\ell}{|\varepsilon|^2}(\rho_{\text{nt}_\ell}-\rho)\right)}
\end{aligned}
\end{eqnarray}
where $\omega$ and $k$=${n_0\omega}/{c}$ are the angular frequency and wavenumber of the carrier wave, $k''$ accounts for the linear group-velocity dispersion, $\rho$ is the electron density, $\rho_{\text{nt}_\ell}$ the density of neutral molecules of species $\ell$, $\rho_c=m_e\epsilon_0\omega/q_e^2$ the critical plasma density and $m_e$ and $q_e$ the mass and charge of the electron, $n_0$ is the linear refractive index at $\lambda$, $W_\ell(|\varepsilon|^2)$ is the photoionization rate of species $\ell$ with ionization potential $U_\ell$, $\sigma_\ell$ is the cross-section for avalanche ionization as defined below in Equation (\ref{avalanche}) and $t$ refers to the retarded time in the reference frame of the pulse. 
The right-hand terms of Eq.(\ref{NLSE}) account for spatial diffraction, second order group-velocity dispersion (GVD), Kerr self-focusing (including the self-steepening term), defocusing by the higher-order nonlinear refractive indices, plasma defocusing, inverse Bremsstrahlung and multiphoton absorption respectively. 
We neglect the delayed orientational response, which for pulses longer than 100~fs would increase the self-focusing term \cite{BergeSNKW07,CouaironM07} and affect the ionization efficiency of  N$_2$ and O$_2$ molecules by less than 20 and 10\%, respectively \cite{LitvinyukLDRVC03,ZhaoTL03}. These opposite effects are negligible in the numerical simulations for 30 fs pulse duration, and do not affect qualitatively the discussion below even for longer pulses. We also neglect space-time focusing. However, in the following we mainly focus on the self-guiding process,  peak intensity and ionization, which are little affected by this approximation.



The spectral dependence of the NLSE (\ref{NLSE}) stems from the plasma contribution as well as the dispersion of the linear ($n_0$) and non-linear ($n_{2j}$) refractive indices. The latter can be deduced at any frequency $\omega$ from those of O$_2$ and N$_2$ at the same frequency by following the Lorenz-Lorentz relation, which links the refractive index of a mix of non-polar gases to its components \cite{Bottcher52}:
\begin{equation}
\frac{n_{mix}^2(\omega)-1}{n_{mix}^2(\omega)+2}=\sum \rho_i \frac{n^2_{i}(\omega)-1}{n^2_{i}(\omega)+2}
\label{LoLo}
\end{equation}
where $n(\omega)$ is defined in Equation (\ref{n_i}) and $\rho_i$ denotes the relative abundance of the species $i$ in the mix. Considering that $\Delta n_{Kerr} \ll 1$, a first-order development of Equation (\ref{LoLo}) yields, for any order $j$:
\begin{equation}
n_{2j,Air}(\omega)=0.79\cdot n_{2j,N_2}(\omega)+0.21 \cdot n_{2j,O_2}(\omega)
\end{equation}

The values of the $n_{2j}$ for argon, nitrogen and oxygen can be calculated at any frequency $\omega$ from their values measured at $\lambda_0$=800 nm \cite{LoriotHFL09} and the dispersion curves in thoses gases \cite{ZhangLW08}, 
through the generalized Miller formula \cite{EttoumiPKW10} expressed in term of non-linear refractive indices:
\begin{equation}
\displaystyle{n_{2j}(\omega) = n_{2j}(\omega_0) \: \left(\frac{n^{2}_{0}(\omega)-1}{n^{2}_{0}(\omega_0)-1}\right)^{2(j+1)}}
\label{miller_n_k}
\end{equation}

While $n_{8,Air}$ varies by a factor of 2 between 300 and 1600 nm, ionization displays an even much steeper spectral dependence. More specifically, the temporal evolution of the electron density is given by

%
\begin{equation}
\label{plasmaeqn}
\displaystyle{\frac{\partial \rho}{\partial t} \approx \sum_{\ell=O_2,N_2} \left(W_\ell(I,\omega)\rho_{0,\text{nt}_\ell} + \frac{\sigma_\ell(\omega)}{U_{\ell}} I \rho\right)}
\end{equation}
where $\rho_{\text{0,nt}_\ell}$ is the initial density of neutral molecules of species $\ell$. Here, attachment to neutral molecules and recombination with positive ions have been neglected owing to the short pulse durations considered in this work. The cross-section for avalanche ionization is calculated on the basis of Drude's theory \cite{BergeSNKW07}:
\begin{eqnarray}
\label{avalanche}
\sigma_\ell(\omega) &=& \frac{q^{2}_\mathrm{e}}{m_{\mathrm{e}} \epsilon_0 n_0(\omega) c \: \nu_{\mathrm{e},\ell} \left(1 + \frac{\omega^2}{\nu_{\mathrm{e},\ell}^2}\right)} \\
&\approx& \frac{q^{2}_{\mathrm{e}} \: \nu_{\mathrm{e},\ell}}{\omega^2 \:m_{\mathrm{e}} \epsilon_0 n_0(\omega) c}
\end{eqnarray}
where $\nu_{\mathrm{e},\ell}$ is the mean collisional frequency of an electron with the species $\ell$ (i.e., the average electron velocity, divided by the mean free path of an electron, assuming that only species $\ell$ is present), and $c$ is the speed of light. At atmospheric pressure in air, $\nu_{\mathrm{e,O_2}}$ = 1/(1.75 ps), $\nu_{\mathrm{e,N_2}}$ = 1/(440 fs), and $\nu_{\mathrm{e,Ar}}$ = 1/(350 fs).
The multiphoton/tunnel ionization rates are given by the multi-species generalized Keldysh-Perelomov, Popov, Terent'ev (PPT) formulation \cite{KasparianSC00,BergeSNKW07}:
\begin{eqnarray}
\begin{aligned}
\label{PPT}
W_\ell(|\varepsilon|^2,\omega) = \frac{4 \sqrt{2}}{\pi} \left|C_{n^{*}_{\ell},l^{*}_{\ell}}\right|^2 \left(\frac{4\sqrt{2} U_\ell^{\frac{3}{2}}}{E_p \sqrt{1+\gamma_\ell^2}}\right)^{2 n^{*}_{\ell}-\frac{3}{2}-\left|m_\ell\right|} \\ \times \: \frac{f(l_\ell,m_\ell)}{|m_\ell|!} \mathrm{e}^{-2 \nu_\ell \left[\mathrm{sinh}^{-1}(\gamma_\ell) - \frac{\gamma_\ell \sqrt{\gamma_\ell^2+1}}{1+2 \gamma_\ell^2}\right]} U_{\ell} \: \frac{\gamma_\ell^2}{1+\gamma_\ell^2} \\ \times \sum^{+\infty}_{\kappa_\ell \geq \nu_{\ell}} \mathrm{e}^{- \alpha_\ell (\kappa_\ell - \nu_\ell)} \: \Phi_m\left(\sqrt{\beta_\ell(\kappa_\ell-\nu_\ell)}\right),
\end{aligned}
\end{eqnarray}
where, expressed in atomic units, $E_p = |\varepsilon| \sqrt{2/(\epsilon_0 c)}$, $\gamma_\ell = \omega \sqrt{2 U_{\ell}}/E_p$, $\nu_\ell = U_{\ell}\left(1+1/(2\gamma^2)\right)/(\hbar \omega)$, $\beta_\ell = 2 \gamma_\ell / \sqrt{1 + \gamma_\ell^2}$, $\alpha_\ell = 2 \left[\mathrm{sinh^{-1}}(\gamma_\ell)-\gamma_\ell/\sqrt{1+\gamma_\ell^2}\right]$, 
and $\Phi_m(x) = \mathrm{e}^{-x^2} \int^{x}_{0} (x^2-y^2)^{|m_{\ell}|}\mathrm{e}^{y^2} \mathrm{d}y$. $n^{*}_{\ell} = Z_\ell/\sqrt{2 U_\ell}$ the effective quantum number, $l^{*}_{\ell} = n^{*}_{\ell}-1$, $l_\ell$ and $m_\ell$ are the orbital momentum and the magnetic quantum number, respectively. In air, $l_\ell=m_\ell=0$ \cite{NuterB06}. $Z_\ell$ is the residual ion charge accounting for the difference between the O$_2$ and N$_2$ molecules and their atomic counterparts. These empirical coefficients $Z_{\text{O}_2}=0.53$ and $Z_{\text{N}_2}=0.9$ have been measured at 800~nm \cite{TalebpourYC99}, and are expected to be independent from wavelength. Since argon is an atomic gas, $Z_{Ar}=1$.
The factors $|C_{n^{*}_{\ell},l^{*}_{\ell}}|$ and $f(l_\ell,m_\ell)$ are
\begin{eqnarray}
    |C_{n^{*}_{\ell},l^{*}_{\ell}}| &=& \frac{2^{2n^{*}_{\ell}}}{n^{*}_{\ell}\Gamma(n^{*}_{\ell} + l^{*}_{\ell} + 1)\Gamma(n^{*}_{\ell}-l^{*}_{\ell})},  \\ [5mm]
    f(l_{\ell},m_{\ell}) &=& \frac{(2l_\ell+1)(l_\ell+|m_\ell|)!}{2^{|m_\ell|} |m_\ell|! (l_\ell - |m_\ell|)!}. 
\end{eqnarray}



To quantify the relative contributions of the higher-order Kerr terms ($n_4$ through $n_{10}$) and of the ionization to the defocusing which balances the self-focusing, we define the instantaneous ratio $\xi$ between the refractive index changes induced by both of these contributions at any location $\vec{r}$:
\begin{equation}
\xi(\vec{r},t)= \lvert \sum_{j\ge2}{n_{2j} I(\vec{r},t)^j}\rvert / \frac{\rho(\vec{r},t)}{2 n_0\rho_c}
\label{xi}
\end{equation}\

This expression is in fact the ratio of the magnitude of the two defocusing terms in the propagation equation (\ref{NLSE}), namely those accounting for higher-order Kerr terms and plasma defocusing, respectively. When considering the overall action of both effects on the whole pulse duration, we define a pulse-integrated value of $\xi$:
\begin{equation}
\overline{\xi}(\vec{r}) ={\int \lvert \sum_{j\ge2}{n_{2j} I(\vec{r},t)^j}\rvert \cdot |\varepsilon(\vec{r},t)| \mathrm{d}t} / {\int \frac{\rho(\vec{r},t)}{2 n_0\rho_c} \: |\varepsilon(\vec{r},t)| \mathrm{d}t},
\end{equation}

\begin{figure}[t!] 
   \begin{center}
      \includegraphics[keepaspectratio, width=12cm]{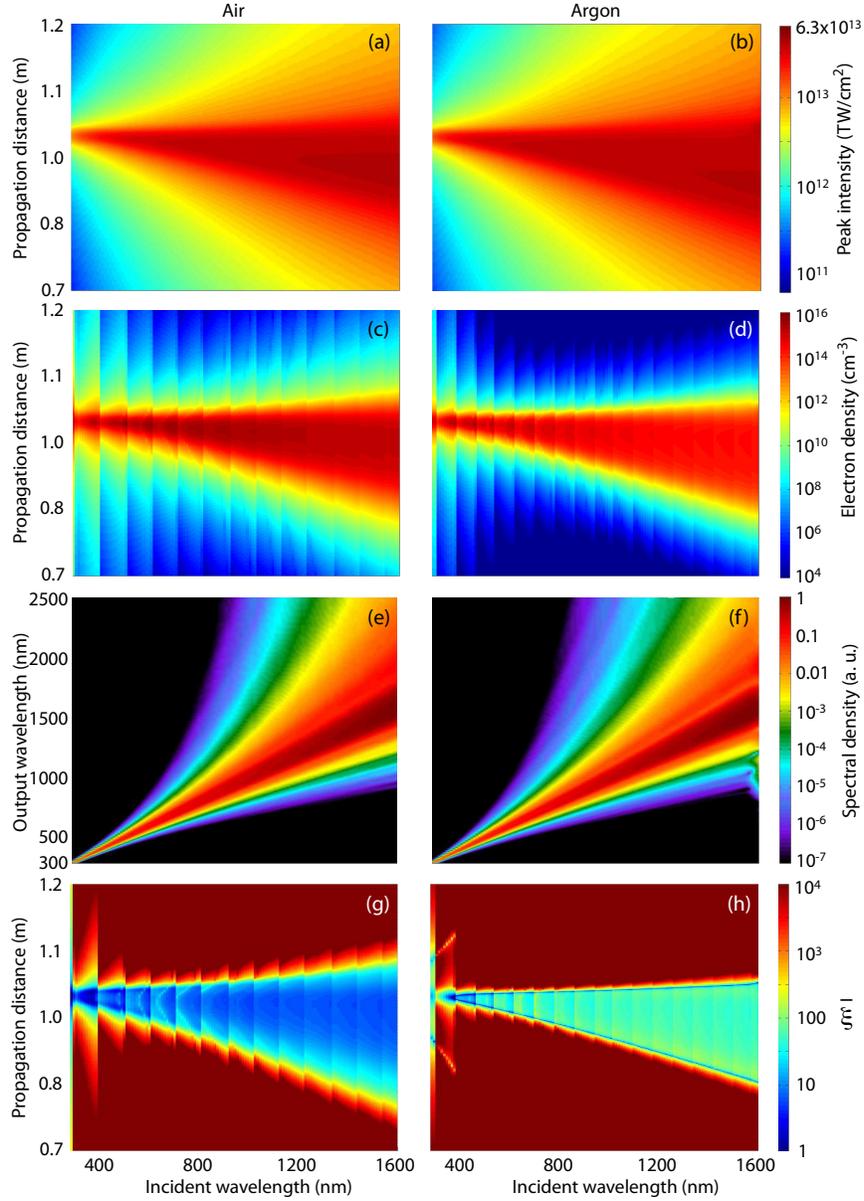}
   \end{center}
   \caption{Maximum on-axis intensity (a,b), electron density (c,d), spectral broadening (e,f) and pulse-averaged ratio $\overline{\xi}$ of the on-axis pulse-averaged contributions of the higher-order Kerr and plasma contributions to the non-linear refractive index (g,h; see text for details), as a function of the wavelength and propagation distance in air (a,c,e,g) and argon (b,d,f,h).}
   \label{intensite_plasma}
\end{figure}

We numerically integrated the propagation equation (\ref{NLSE}) for an ultrashort pulse typical of laboratory-scale filamentation experiments: $30$~fs Fourier-limited FWHM pulse duration, a peak power of 6.5 critical powers $P_{\mbox{cr}}=\lambda^2/4\pi n_2$, a beam diameter of $\sigma_{\mbox{r}} = 4$ mm, a focal length $f=1$ m and a pressure of $1$ bar of either air or argon. The main results are displayed in Figure \ref{intensite_plasma}.

Our simulations yield filamentation over the whole investigated spectral range (300 - 1600 nm), reproducing experimental observations from the UV \cite{SchwarzRDKWM00,TzortzakisLCFPM00} to the mid-IR \cite{HauriLBSCCCDMRSTWDP07}.
As is visible in Figure \ref{intensite_plasma} (a) and (b), the clamping intensity, filament onset and filament length are very similar in air and argon over the whole considered spectral length for a given incident reduced power $P/P_{cr}$. Moreover, longer wavelengths yield longer filaments with an earlier onset. Furthermore, the output spectrum after 1.5 m propagation (panels (e) and (f)) is broader for longer wavelengths. As a result, the relative broadening, defined as the ratio of the output spectral width to the initial frequency, $\Delta\omega/\omega_0$, is almost constant across the spectrum. Spectral properties appear very similar in air and argon. 

\begin{figure}[tb!]
   \begin{center}
       \includegraphics[keepaspectratio, width=8cm]{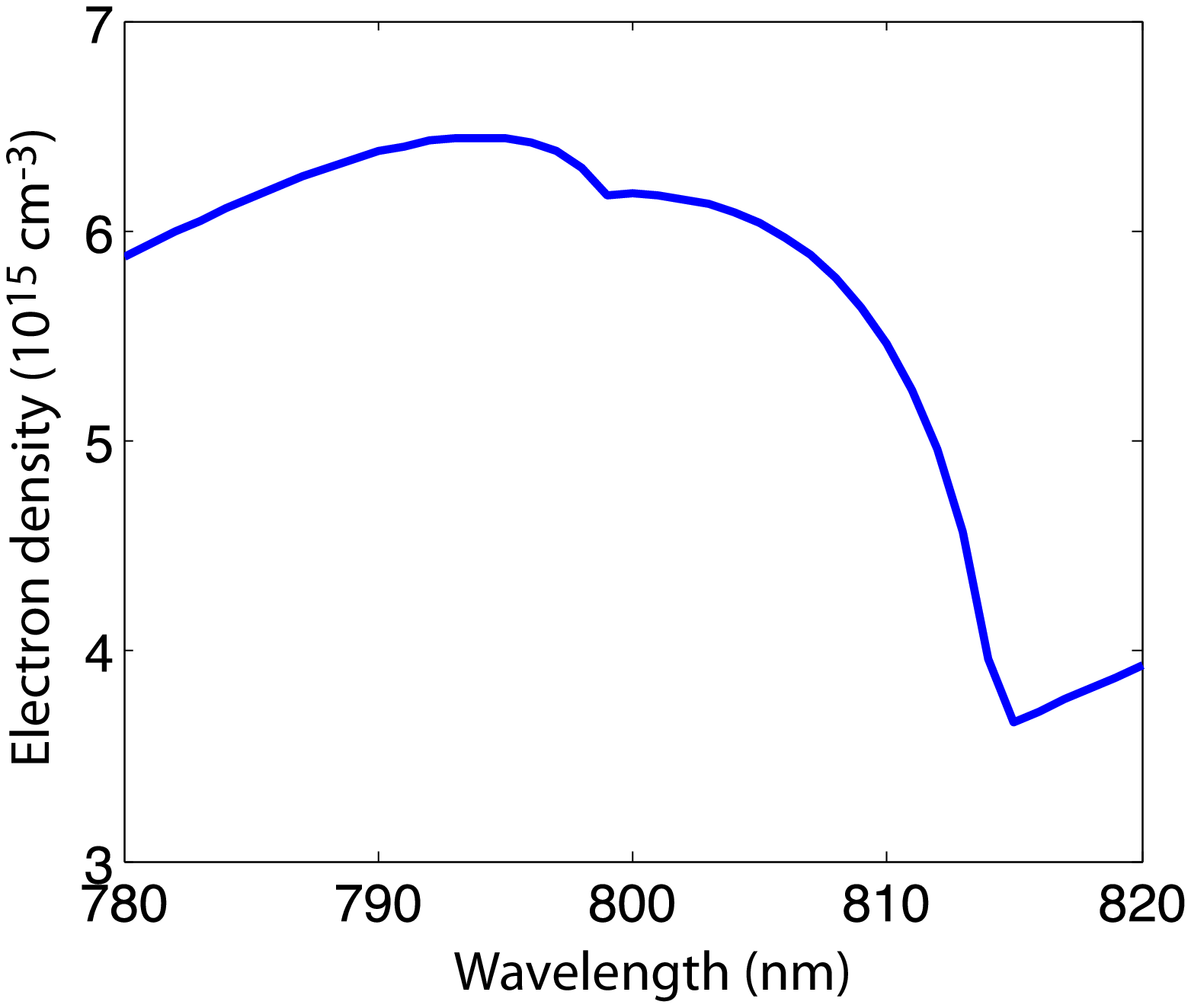}
   \end{center}
   \caption{Wavelength dependence close to 800 nm of the electron density generated by a 30 fs pulse of constant intensity 50 TW}
   \label{contributions}
\end{figure}

\begin{figure}[tb!]
   \begin{center}
       \includegraphics[keepaspectratio, width=8.5cm]{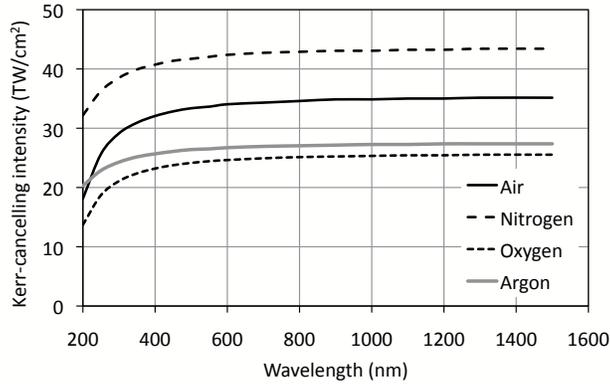}
   \end{center}
   \caption{Wavelength dependence of the clamping intensity canceling the Kerr effect $I_{\Delta n_\text{Kerr}=0}$, in O$_2$, N$_2$, air and Ar, based on the dispersion of the higher-order Kerr indices from Equation (\ref{miller_n_k})}
   \label{I_delta_n_kerr}
\end{figure}

\begin{figure}[t!] 
   \begin{center}
      \includegraphics[keepaspectratio, width=12cm]{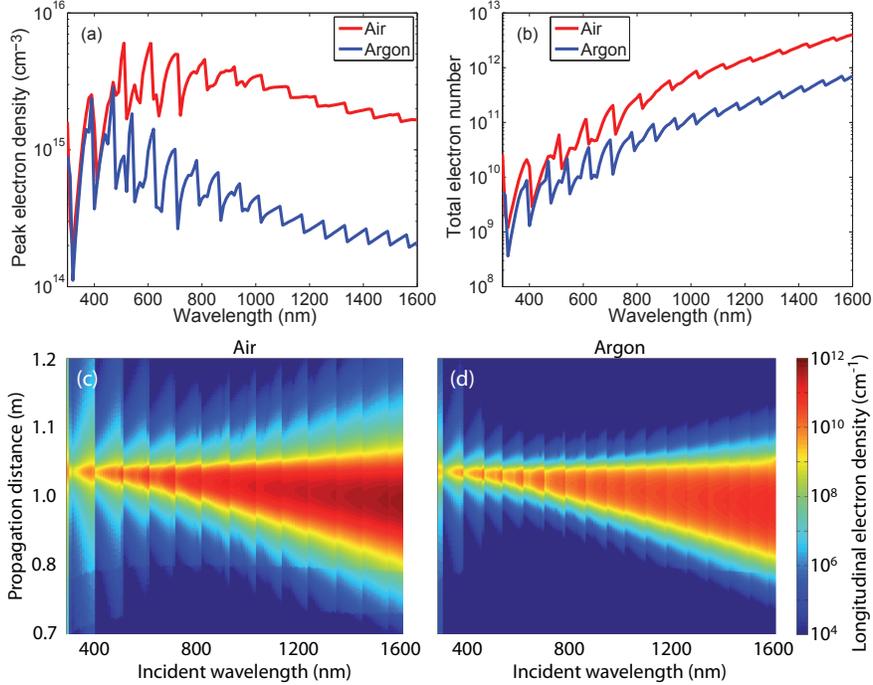}
   \end{center}
   \caption{Wavelength dependence of ionization in laser filamentation. (a) Peak electron density; (b) total number of electrons; (c,d) Transverse-integrated longitudinal electron density in air and argon, representative of the electric conductivity of the filament. Note the difference with Figures \ref{intensite_plasma}(c,d), which is due to the larger filament diameter in the infrared.}
   \label{plasma_total}
\end{figure}

In contrast, the electron density is approximately 10 times higher in the former than in the latter. The spectral dependence of the electron density is very non-monotonic (Panels (c) and (d)). On one hand, due to steps in the number of photons required for ionization,  the electron density is not a monotonic function of the wavelength. For instance, ionization in air is almost 2 times more efficient at 793 nm than at 815 nm (See Figure \ref{contributions}). On the other hand, contrary to expectations that shorter wavelengths should result in stronger ionization, a maximum in the peak electron density is observed in the 610~nm range in the case of air, and around 470~nm in argon (Figures \ref{intensite_plasma}(c),(d) and \ref{plasma_total}(a)). Away from this maximum, the peak electron density decreases within a dynamics of typically one order of magnitude. This unexpected behavior stems from the convolution of the respective spectral dependences of (i) the ionization rate $W$, which increases at shorter wavelengths and (ii) the peak intensity, which decreases in the UV. 
This decrease is due to the fact that, according to Equation (\ref{miller_n_k}) and considering the typical dispersion curves in gases, the higher-order indices increase faster, in absolute values, when the frequency increases. As a consequence, the clamping intensity canceling the Kerr effect $I_{\Delta n_\text{Kerr}=0}$ is lower, as shown on Figure \ref{I_delta_n_kerr}.

The spectral dependence of the peak electron density by more than one order of magnitude, also evidenced on Figure \ref{plasma_total}(a), appears in contradiction with numerical results obtained without considering the higher-order Kerr terms \cite{SkupinB07}, which predict almost similar electron densities, close to $10^{16}$ cm$^{-3}$, at the three investigated wavelengths (248, 800 and 1550 nm). Since the latter results were obtained with longer pulses (127 fs) of slightly lower power than in our work, a direct quantitative comparison cannot be performed. However, we expect that the qualitative difference between our results and those of Ref. \cite{SkupinB07} are due to the inclusion of the higher-order Kerr terms in our model, and thus illustrate the influence of the latter.

 
The very similar behavior, except for the plasma density, of two gases with comparable values of the non-linear refractive indices but significantly different ionization potentials, suggest that filamentation is driven by the higher-order Kerr terms rather than by plasma, not only at 800~nm as evidenced recently \cite{LoriotBHFLHKW09}, but also on the whole investigated spectral range in both air and argon. Indeed, the on-axis $\overline{\xi}$ ratio (Figure \ref{intensite_plasma}(g),(h)) exceeds 1 in all considered conditions. Furthermore, the higher-order Kerr terms strongly dominate the filamentation dynamics ($\overline{\xi}>10$) at all wavelengths above $\lambda\sim$400~nm in argon and above $\lambda\sim$600~nm in air, although, due to the thresholds induced by the steps in photon numbers required to ionize oxygen (See Equation (\ref{PPT})), the evolution of this behavior is not strictly monotonic. Due to the domination of the higher-order Kerr terms, the latter will govern the intensity within the filaments, so that the ionization is mostly decoupled from the filamentation dynamics.


It should be noted that Miller formul\ae\ are sometimes considered to underestimate dispersion \cite{MizrahiS85}. However, such correction to the dispersion curve of higher-order Kerr terms would result in a larger absolute value of the high-order Kerr indices in the UV. In this case, $I_{\Delta n_\text{Kerr}=0}$ will be even lower in the UV. In other words, the curves of Figure \ref{I_delta_n_kerr} will decrease faster on the left side of the graph. As a consequence,  the higher-order Kerr indices will have an even more dominant contribution to the self-guiding, as compared to that of plasma. Such larger dispersion would therefore reinforce qualitatively our conclusion about the domination of defocusing higher-order Kerr terms over the plasma defocusing. 

This new vision of filamentation provides straightforward interpretation to experimental observations that the spectral broadening mostly occurs at the beginning of the filaments, while the spectrum is little affected in the main region of self-guided propagation \cite{Mondelain01}. This is due to the fact that, in the self-guided propagation, the intensity is clamped close to  $I_{\Delta n_\text{Kerr}=0}$, so that the spectral counterpart of this Kerr effect, self-phase modulation (SPM), is also waved out. The earlier onset of filamentation for longer wavelengths is compatible with the Marburger formula \cite{Marburger}, which predicts that the non-linear focal length describing self-focusing is inversely proportional to the wavelength. Furthermore, the longer filaments in the infrared appears to stem from the lower peak electron density (Figure \ref{plasma_total}(a)).

The very smooth dispersion curve of both air and argon above 500~nm \cite{EttoumiPKW10} results in a very slow variation of this clamping intensity $I_{\Delta n_\text{Kerr}=0}$ (Figure \ref{I_delta_n_kerr}) and explains the quasi-constant clamping intensity observed over this spectral range. 

The availability, for the first time, of simulations at constant reduced power $P/P_{cr}$ over a broad spectral range allows to discuss the choice of wavelength to optimize the filaments properties for specific purposes. As described above, the spectral dependence of the electron generation contrasts strongly with the common expectation that, due to a more efficient ionization, ultraviolet wavelengths should ionize the propagation medium more efficiently. 

However, the peak electron density is not directly relevant for typical atmospheric applications. For example, laser-assisted water condensation \cite{RohwetterKSHLNPQSSHWW09}, can be expected to require the largest possible value of the total generated charge. This total charge is obtained by integrating the plasma density over the filament length and cross-section (Figure \ref{plasma_total}(b)). Here, the longer filaments in the infrared as well as their larger diameter result in an unanticipated larger total charge at longer wavelengths.

Applications such as lightning control \cite{KasparianAAMMPRSSYMSWW08} require the longest possible filaments with the higher conductivity, i.e. a high longitudinal (transversely integrated) electron density. Long wavelengths simultaneously optimize both of these properties (Figure \ref{plasma_total}(c,d)).  In that regard, UV lasers providing short filaments with a small diameter and decreasing transverse-integrated electron densities would expectedly be less efficient than the commonly used Ti:Sa lasers around 800 nm, while wavelengths further in the IR should be even more efficient.
Infrared is also very attractive because it would alllow to work in the "telecom" spectral region (1.55 $\mu$m), where optical components are available and eye-safety standards are much more favorable for atmospheric applications than in the rest of the NIR and visible spectral region.

In conclusion, applying generalized Miller formul\ae~to estimate arbitrary-order non-linearities from the infrared to the ultraviolet spectrum, we have shown that higher-order nonlinear indices (up to $n_8$ in air and $n_{10}$ in argon) have a dominant contribution to both the focusing and defocusing terms implied in the self-guiding of 30~fs laser pulses in air, at all wavelengths between 300 and 1600~nm. As a consequence, the plasma generation is almost decoupled from the self-guiding of filaments over the whole visible and infrared spectral range. Instead, filaments can be considered as a self-guiding regime mostly governed in air and argon by the balance between the alternate signs of the nonlinear indices, respectively resulting in Kerr self-focusing and self-defocusing. 

Moreover, a systematic investigation at constant reduced power $P/P_{cr}$ as a function of wavelength provides for the first time hints to choose optimal wavelengths for generating filaments optimized for a specific applications. In particular, we have shown that, contrary to previous expectations, the infrared is more favorable than the commonly used 800 nm wavelength if long ionized filaments or high total amount of charges are required. This raises the hope to further improve the spectacular results \cite{KasparianW08,KasparianAAMMPRSSYMSWW08,RohwetterKSHLNPQSSHWW09} recently obtained with Titanium:Sapphire lasers. Moreover, the higher eye-safety threshold in the telecom window at 1.55 $\mu$m is favorable for the practical development of the evisioned applications in free space.

Acknowledgements. This work was supported by the Conseil R\'egional
de Bourgogne, the ANR \textit{COMOC}, the \textit{FASTQUAST} ITN Program of the 7th FP and the Swiss NSF (contracts 200021-116198 and 200021-125315).
\bibliographystyle{unsrt}

\begin{thebibliography}{10}

\bibitem{BraunKLDSM95}
A.~Braun, G.~Korn, X.~Liu, D.~Du, J.~Squier, and G.~Mourou.
\newblock {Opt. Lett.} \textbf{20}, 73 (1995).

\bibitem{ChinHLLTABKKS05}
S.~L. Chin, S.~A. Hosseini, W.~Liu, Q.~Luo, F.~Theberge, N.~Ak\"ozbek, A.~Becker, V.~P. Kandidov, O.~G. Kosareva, and H.~Schroeder.
\newblock {Can. J. Phys.} \textbf{83}, 863 (2005).

\bibitem{BergeSNKW07}
L.~Berg\'e, S.~Skupin, R.~Nuter, J.~Kasparian, and J.-P.~Wolf.
\newblock {Rep. Prog. Phys.} \textbf{70}, 1633 (2007).

\bibitem{CouaironM07}
A.~Couairon and A.~Mysyrowicz.
\newblock {Phys. Rep.}, \textbf{441} 47 (2007).

\bibitem{KasparianW08}
J.~Kasparian and J.-P. Wolf.
\newblock {Opt. Express} \textbf{16}, 466 (2008).

\bibitem{AkozbekSBC01}
N.~Ak\"ozbek, M.~Scalora, C.M. Bowden, and S.L. Chin.
\newblock {Opt. Comm.} \textbf{191}, 353 (2001).

\bibitem{VincotteB04}
A.~Vin\c{c}otte and L.~Berg\'e,
\newblock {Phys. Rev. A} \textbf{70}, 061802(R) (2004).

\bibitem{FibichI04}
G. Fibich and B. Ilan, 
Opt. Lett. \textbf{29}, 887 (2004)

\bibitem{BergeSMKYFSW05}
L. Berg\'e, S. Skupin, G. M\'ejean, J. Kasparian, J. Yu, S. Frey, E. Salmon, and J.P. Wolf, 
Phys. Rev. E \textbf{71}, 016602(2005)

\bibitem{ZhangTWDW10}
J.-F. Zhang, Q. Tian, Y.-Y. Wang, C.-Q. Dai, L. Wu,
Phys. Rev. A \textbf{81}, 023832 (2010)

\bibitem{LoriotHFL09}
V.~Loriot, E.~Hertz, O.~Faucher, B.~Lavorel,
Opt. Express \textbf{16}, 13429 (2009); Erratum in Opt. Express \textbf{18} 3011 (2010)

\bibitem{LoriotBHFLHKW09}
P. B\'ejot, J. Kasparian, S. Henin, V. Loriot, T. Vieillard, E. Hertz, O. Faucher, B. Lavorel, and J.-P. Wolf, 
Phys. Rev. Lett. \textbf{104}, 103903 (2010)

\bibitem{MechainCADFPTMS04}
G.~M\'echain, \emph{et al.}, 
\newblock {Appl. Phys. B} \textbf{79}, 379 (2004).

\bibitem{DubietisGTT04}
A. Dubietis, E. Gaizauskas, G. Tamosauskas, and P.Di Trapani, 
Phys. Rev. Lett. \textbf{92}, 253903 (2004)

\bibitem{StibenzZS06}
G.~Stibenz, N.~Zhavoronkov, G.~Steinmeyer.
\newblock {Opt. Lett.} \textbf{31}, 274 (2006)

\bibitem{KasparianSC00}
J.~Kasparian, R.~Sauerbrey, and S.~L. Chin.
\newblock {Appl. Phys. B} \textbf{71}, 877 (2000).

\bibitem{ThebergeLSBC06}
F. Th\'eberge, W. Liu, P. Tr. Simard, A. Becker, and S. L. Chin,
Phys. Rev. E \textbf{74}, 036406 (2006)

\bibitem{SchmidtBGSTBKWVKCL10}
B.E. Schmidt, P. B\'ejot, M. Gigu\`ere, A.D. Shiner, C. Trallero-Herrero, E. Bisson, J. Kasparian, J.-P. Wolf, D.M. Villeneuve, J.-C. Kieffer, P.B. Corkum, and F. L\'egar\'e, 
Applied Physics Letters. \textbf{96}, 121109 (2010)

\bibitem{BejotSKWL10}
P. B\'ejot, B. E. Schmidt, J. Kasparian, J.-P. Wolf, F. Legar\'e,
Phys. Rev. A \textbf{81}, 063828 (2010)

\bibitem{SchwartzRD00}
J. Schwarz, P. Rambo, and J.-C. Diels, 
Optics Communications. \textbf{180}, 383 (2000)

\bibitem{EttoumiPKW10}
W. Ettoumi, Y. Petit, J. Kasparian, J.-P. Wolf, Opt. Express \textbf{18}, 6613 (2010)

\bibitem{LitvinyukLDRVC03}
I.V. Litvinyuk, Kevin F. Lee, P.W. Dooley, D.M. Rayner, D.M. Villeneuve, and P. B. Corkum,
Phys. Rev. Lett. \textbf{90} 233003 (2003)

\bibitem{ZhaoTL03}
Z. X. Zhao, X. M. Tong, C. D. Lin, 
Phys. Rev. A \textbf{67}, 043404 (2003)

\bibitem{Bottcher52}
C. J. F. B\"ottcher, \emph{Theory of Electric Polarisation}, Elsevier Publishing Company, Amsterdam, 1952

\bibitem{ZhangLW08}
J. Zhang, Z. H. Lu, L. J. Wang,
Appl. Opt. \textbf{47}, 3143 (2008)



\bibitem{NuterB06}
R. Nuter and L. Berg\'e, J. Opt. Soc. Am. B \textbf{23}, 874 (2006)

\bibitem{TalebpourYC99}
A. Talebpour, J. Yang, S. L. Chin, Opt. Commun. \textbf{163}, 29 (1999)

\bibitem{SchwarzRDKWM00}
J. Schwarz, P. Rambo, J.-C. Diels, M. Kolesik, E. M. Wright, J. V. Moloney,
Optics Comm. \textbf{180}, 383 (2000) 

\bibitem{TzortzakisLCFPM00}
S. Tzortzakis, B. Lamouroux, A. Chiron, M. Franco, B. Prade, A. Mysyrowicz, S. D. Moustaizis
Opt. Lett. \textbf{25}, 1270 (2000)

\bibitem{HauriLBSCCCDMRSTWDP07}
C. P. Hauri, R. B. Lopez-Martens, C. I. Blaga, K. D. Schultz, J. Cryan, R. Chirla, P. Colosimo, G. Doumy, A. M. March, C. Roedig, E. Sistrunk, J. Tate, J. Wheeler, L. F. DiMauro, E. P. Power,
Opt. Lett. \textbf{32}, 868 (2007)

\bibitem{SkupinB07}
S. Skupin, L. Berg\'e
Opt. Comm. \textbf{280}, 173 (2007)


\bibitem{MizrahiS85}
V. Mizrahi, D.P. Shelton, 
Phys. Rev. Lett. \textbf{55}, 696 (1985)

\bibitem{Mondelain01}
D. Mondelain, 
Ph. D. thesis, Universit\'e Lyon I, 2001. http://tel.archives-ouvertes.fr/tel-00396346/fr/


\bibitem{Marburger}
E.L. Dawes and J.H. Marburger, 
Phys. Rev. \textbf{179}, 862 (1969)



\bibitem{RohwetterKSHLNPQSSHWW09}
P. Rohwetter, J. Kasparian, K. Stelmaszczyk, S. Henin, N. Lascoux, W. M. Nakaema, Y. Petit, M. Quei\ss er, R. Salam\'e, E. Salmon, Z.Q. Hao, L. W\"oste, J.-P. Wolf, 
Nature Photonics, \textbf{4}, 451 (2010) 

\bibitem{KasparianAAMMPRSSYMSWW08}
J. Kasparian, R. Ackermann, Y.-B. Andr\'e, G. M\'echain, G. M\'ejean, B. Prade, P. Rohwetter, E. Salmon, K. Stelmaszczyk, J. Yu, A. Mysyrowicz, R. Sauerbrey, L. W\"oste, and J.-P. Wolf, 
Optics Express. \textbf{16}, 5757 (2008)

\end{thebibliography}


\end{document}